# Design and Mathematical Modelling of Inter Spike Interval of Temporal Neuromorphic Encoder for Image Recognition

Aadhitiya VS[1], Jani Babu Shaik[2], Sonal Singhal[3]
[1,2,3]*Department of Electrical Engineering*
*Shiv Nadar University*
Greater Noida, India
[1]av558@snu.edu.in, [2]skjanibabu786@gmail.com,
[3]sonal.singhal@snu.edu.in

Siona Menezes Picardo[4], Nilesh Goel[5]
[4,5]*Department Electrical and Electronics Engineering*
*BITS Pilani Dubai Campus*
Dubai, UAE
[4]sionacmenezes@gmail.com, [5]goel.nilesh@gmail.com

***Abstract*— Neuromorphic computing systems emulate the electrophysiological behavior of the biological nervous system using mixed-mode analog or digital VLSI circuits. These systems show superior accuracy and power efficiency in carrying out cognitive tasks. The neural network architecture used in neuromorphic computing systems is spiking neural networks (SNNs) analogous to the biological nervous system. SNN operates on spike trains as a function of time. A neuromorphic encoder converts sensory data into spike trains. In this paper, a low-power neuromorphic encoder for image processing is implemented. A mathematical model between pixels of an image and the inter-spike intervals is also formulated. Wherein an exponential relationship between pixels and inter-spike intervals is obtained. Finally, the mathematical equation is validated with circuit simulation.**

## I. INTRODUCTION

The term neuromorphic computing was coined by Carver Mead in 1990 [1]. It aims to parallel the electrophysiological behavior of the neuronal systems in the brain using VLSI circuits. Neuromorphic computing devices and circuits have gained remarkable attention in recent times [2][3]. It is mainly due to their ability to carry out cognitive tasks like image recognition with superior accuracy in the low power regime [4]. Research on the hardware implementation of machine learning algorithms for image recognition has been substantial in the past few years. Since neuronal systems operate on spike trains, the encoding of sensory data into spike trains is necessary for neuromorphic devices to function accurately.

In general, encoding is the process of transforming signal information from one form of representation to another. In neuromorphic systems, the input sensory data is transmitted through spike trains. Three neuromorphic-based encoding techniques have mainly been reported in the literature [5]–[8]; (i) time to first spike (TFS) encoding, (ii) rate encoding, and (iii) temporal encoding. TFS has been studied and implemented in [5], [7], [8]. However, it carries information in a single dimension, limiting its versatility and flexibility [8]. The rate encoding technique is extensively used in hardware applications due to its simplicity. However, rate encoding does not genuinely mimic the behavior of neurons, and there is a loss of temporal information [8].

On the other hand, temporal encoding maps information into the relative time between the subsequent spikes rather than the absolute time. This engenders the ability to map higher dimensional information of the original signal [8], [9]. Therefore, temporal encoding can be used in reservoir computing systems.

Fig.1 illustrates the conversion of an image into spike trains. Encoding of an image involves the conversion of the pixels into spike trains. First, the pixel value is converted to a voltage value and is fed as the input to the encoder. Next, these voltage values are fed to all the neurons in the neural network's input layer to generate spikes.

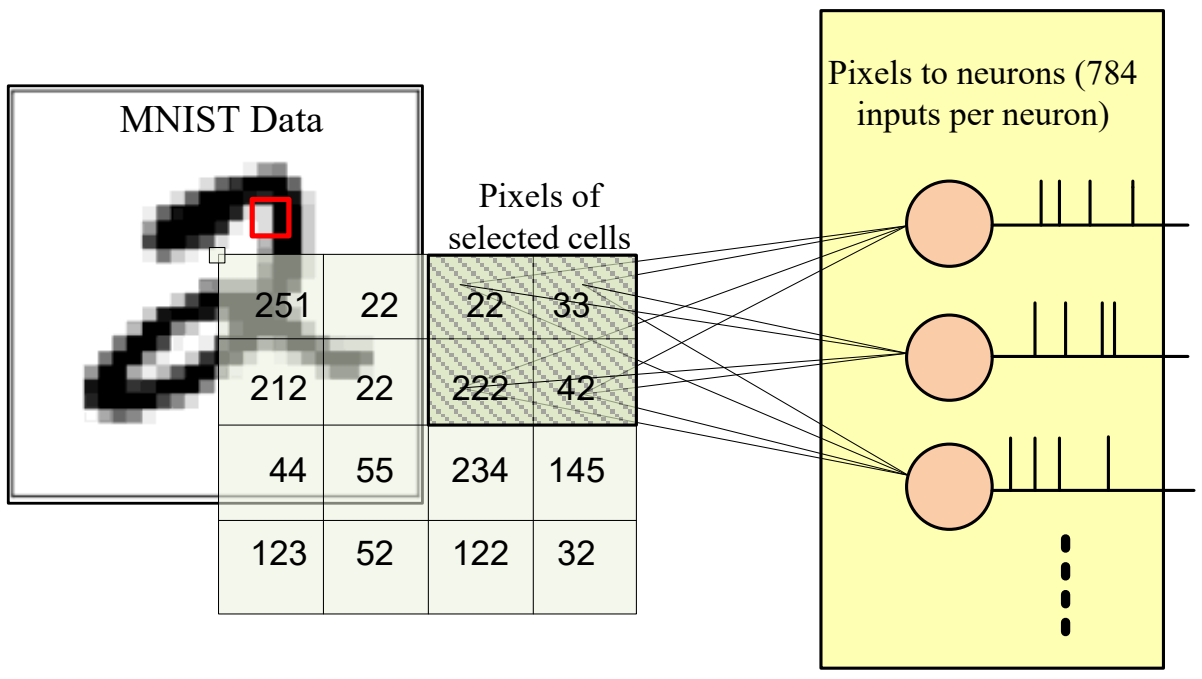

Fig. 1. Neuromorphic spike encoding mechanism for an input image

With a plethora of image recognition applications and the superiority of the temporal encoding scheme in mimicking neural activity, a mathematical model and design could ease the ease of understanding the conversion of pixels into spike trains. In this work, we implement a low-power neuromorphic encoder and formulate a mathematical model between the pixels of an image and the inter-spike intervals.

The remaining paper is organized as follows: Section 2 describes the design and the mathematical modeling of the neuromorphic temporal encoder. Then, results and discussions are presented in section 3. Finally, section 4 concludes this work.

## II. DESIGN AND MATHEMATICAL MODELLING

In this section, the design and the mathematical model of the temporal encoder are presented and discussed.

Fig. 2 shows the block diagram of the temporal neuromorphic encoder. The circuit contains four modules: (i) pixel to excitatory current converter, (ii) leaky transistors & membrane capacitors, (iii) voltage bias generator, and (iv) neurons. All circuits are implemented on industry-standard HKMG based 45nm technology.

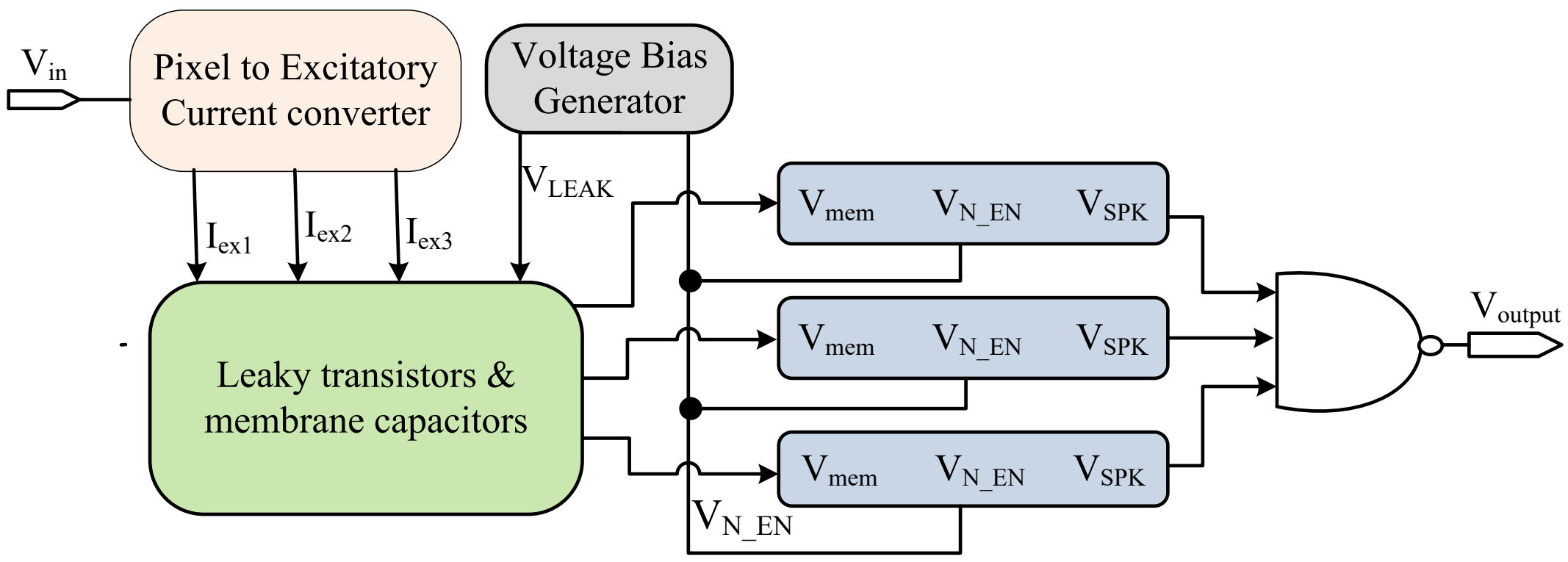

Fig. 2. Block diagram of the temporal neuromorphic encoder

Image pixel intensity ranging from 0 to 255 is represented by input voltage ($V_{in}$). Here, in this work, the $V_{in}$ and image pixel and are related by equation (1):

$$V_{in}(mv) = \frac{pixel}{2} \quad (1)$$

Fig. 3 shows the schematic of the pixel to excitatory current converter module. For low-power operation, transistors in this module are held in the sub-threshold regime. First, the input voltage ($V_{in}$) converts into the input current ($I_{in}$) depending on the gate-source voltage of PM0. Then, $I_{in}$ is mirrored as the excitatory current ($I_{ex}$) using the current mirror circuits. The excitatory current magnitude can be varied by adjusting the dimensions of the weighted transistors. The second module, the voltage bias generator, generates a constant leak voltage ($V_{leak}$) of approximately 250mV. The $V_{leak}$ and weighted excitatory currents ($I_{ex}$) are fed to the third module, i.e., membrane capacitors and leaky transistors. The excitatory current develops membrane potential ($V_{mem}$) across membrane capacitor ($C_{mem}$). These $C_{mem}$ and leaky transistors are part of the neuron circuit. A simplified leaky integrate-and-fire (SLIF) neuron circuit is considered in this work [10]. The neuron circuit will generate a spike-event as developed $V_{mem}$ reaches the membrane threshold ($V_{TM}$).

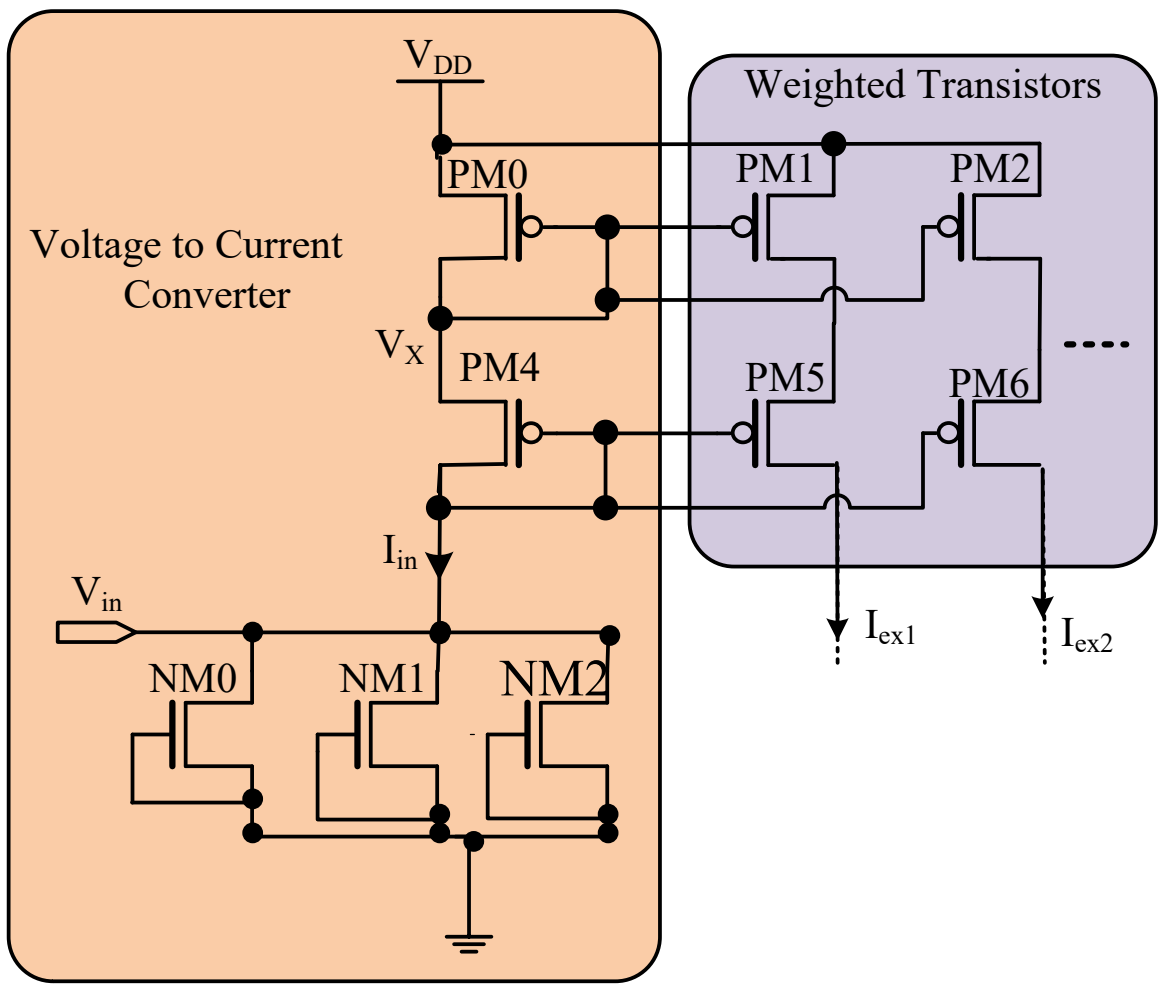

Fig. 3. Schematic of pixel to the excitatory current module

Pixels (0-255) are mapped to input voltages, which ensure excitatory currents, are well within the limits of the neuron. This forms the basis for the mathematical relationship between pixels and inter-spike interval.

As discussed earlier, transistor PM0 is held in the subthreshold region of operation, which aids in minimizing the current magnitude, thereby making the encoder more power-efficient. PM4 is operated in the subthreshold region and is cascaded with PM0, ensuring minimal voltage swing in the drain to source voltage ($V_{DS}$) of PM0. Thus, the input current ($I_{in}$) is observed to show an exponential dependence with $V_{in}$ and is modeled using equation (2):

$$I_{in} = I_0 e^{\frac{V_X - V_{in} - |V_{tp}|}{s.U_T}} \quad (2)$$

$V_X$ is the intermediate potential, $V_{tp}$ is the threshold voltage of PM4, s is the process parameter, $U_T$ is the thermal voltage. Since PM0 and PM4 are diode-connected transistors, the voltage at $V_X$ is the average of $V_{DD}$ and $V_{in}$. $V_X$ can be written by equation (3).

$$V_X = \frac{V_{DD} + V_{IN}}{2} \quad (3)$$

Due to a current mirror mechanism between PM0 and PM1, the current PM1 ($I_{ex1}$) is the mirrored current of PM0 ($I_{in}$). It is given by the following equation (4).

$$I_{ex} = K * e^{\frac{V_X - V_{in} - |V_{tp}|}{s.U_T}} \quad (4)$$

Where K is a weight value that is fixed by transistor dimensions. Substituting (3) in (4), we get equation (5).

$$I_{ex} = Ke^{\frac{\frac{V_{DD} - V_{in}}{2} - |V_{tp}|}{s.U_T}} \quad (5)$$

Equation (5) reports the relationship between $I_{ex}$ and $V_{in}$. This relationship can be used to model the conversion of sensory data into the excitatory current.

Substituting (1) in (5), we get the relationship between $I_{ex}$ and pixels.

$$I_{ex} = K * e^{\frac{2V_{DD} - pixel - 4|V_{tp}|}{4s * U_T}} \quad (6)$$

For image processing applications in general, the sampling speed for encoding images and the resolution of mapping images are essential parameters. The sampling speed for processing images in the neuromorphic encoder is governed by an external voltage source ($V_{N_EN}$). The considered sampling speed in this work is 1.1 MHz. The resolution of mapping images here refers to the accommodation of pixel intensities (0-255) within the range of operating input voltage. It is the difference between two subsequent pixel values expressed in terms of input voltage $V_{in}$. In the temporal encoder, these two factors have a complementary effect on each other. The mapping should be done in such a way that all the pixels are converted to a range of voltage values for which an equal number of spike trains are generated. There could be a loss of information and might lead to erroneous inter-spike intervals. Thus, for faster encoding, the resolution of the mapping needs to be compromised. Higher the difference, the more the resolution. By doing this, a more significant number of pixels will be represented in a smaller range of voltage values, making the inter-spike intervals challenging to interpret because of their minimal difference. The operating region of the encoder needs to be determined to optimize the resolution for a given speed. The time taken to encode an image of M x N pixels is $M*N*T_{samp}$, where $T_{samp}$ is the period of the neuron enable ($V_{N_EN}$).

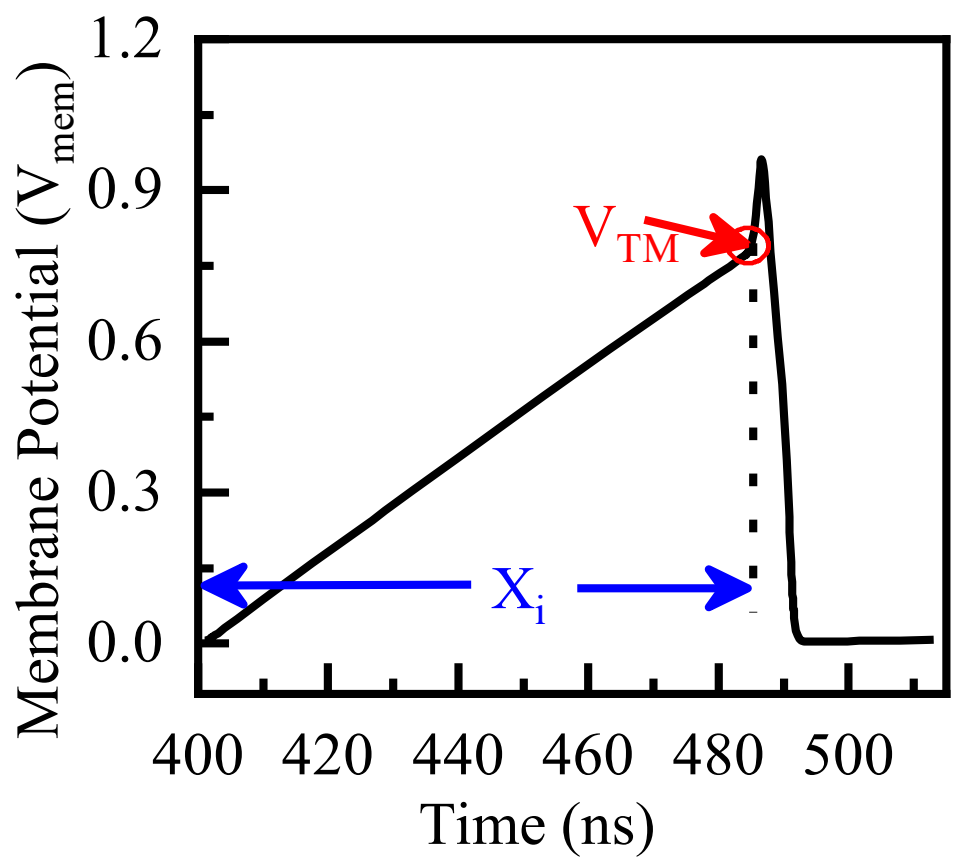

Fig. 4. The simulated waveform of membrane potential ($V_{mem}$)

Fig. 4 shows the simulated waveform of the membrane potential ($V_{mem}$) of the SLIF neuron circuit. The membrane threshold voltage ($V_{TM}$) is highlighted in the figure. The integrating time is the time for $V_{mem}$ to reach the $V_{TM}$, as shown in the figure. It is represented as $X_i$ for the $i^{th}$ SLIF neuron. In reference [8], the integrating time for the SLIF neuron is calculated using equation (7).

$$X_i = \frac{C_{memi} * V_{TM}}{I_{exi} - I_{leak}} \tag{7}$$

Where $I_{exi}$, $I_{leak}$, $C_{memi}$, and $V_{TM}$, are $i^{th}$ branch excitatory current, leak current, membrane capacitor, and membrane threshold voltage, respectively. $I_{leak}$ is negligible compared to excitatory current $I_{ex}$. Thus, omitting $I_{leak}$ current for further analysis.

Fig. 5 shows the timing diagram of encoder output for a sampled period of neuron enable signal ($V_{N_EN}$). The inter-spike interval is the time difference between two subsequent spikes, as shown in the figure. The $i^{th}$ inter-spike interval ($D_i$) can be calculated by equation (8).

$$D_i = X_{i+1} - X_i \tag{8}$$

Where $X_i$ is the $i^{th}$ neuron integrating time to spike (marked in Fig. 4).

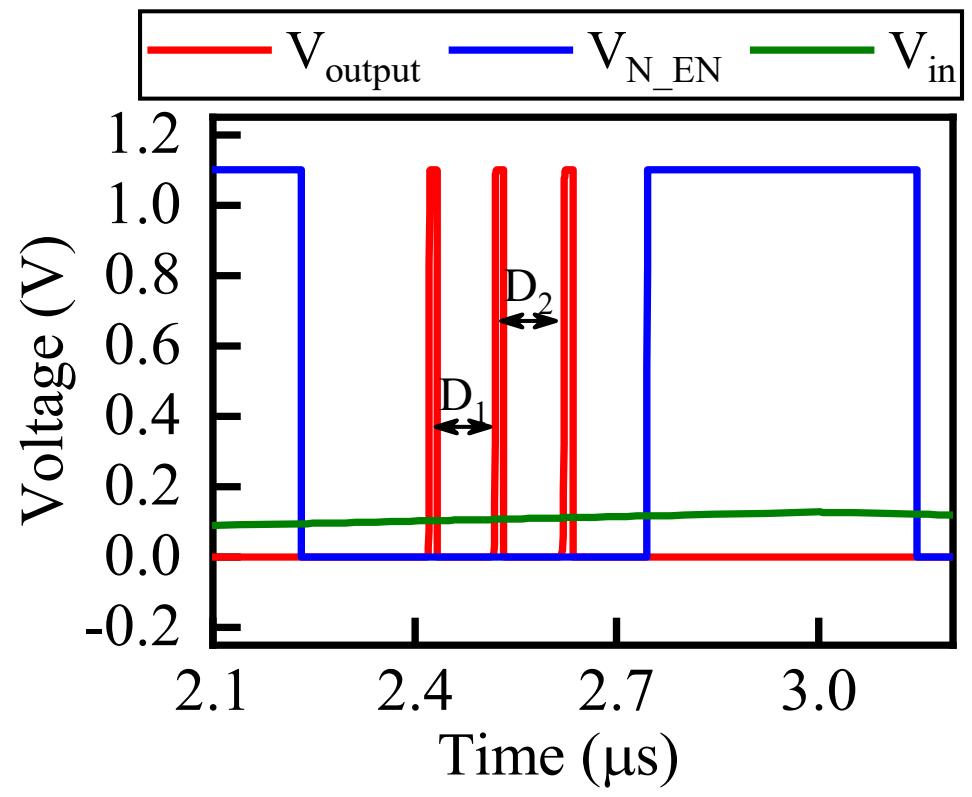

Fig. 5. Illustrating the inter-spike intervals ($D_1$ and $D_2$) for input voltage ($V_{in}$). $V_{N_EN}$ is sampled clock signal

Finally, the relationship between $D_1$ and $D_2$ with pixels is computed by substituting equations (6) & (7) in (8). Thus, obtaining an exponential relation between pixel and the inter-spike interval is given by equation (9).

$$D_i = (C_{memi+1} - C_{memi}) * V_{TM} * K_1 . e^{\frac{(pixel + 4|V_{tp}| - 2V_{DD})}{4\,s.U_T}} \tag{9}$$

Where $C_{memi+1}$ and $C_{memi}$ are the $i+1^{th}$ and $i^{th}$ branch membrane capacitance, $K_1$ is the inverse of the device process parameter dependent on the dimensions of weighted transistors.

## III. Results and Discussions

This section discusses the circuit simulation results of inter-spike intervals ($D_1$ and $D_2$) with image pixels. Further, $D_1$ and $D_2$ have been calibrated from equation (9) are compared with the inter-spike intervals from the circuit simulation to analyze the encoder accuracy.

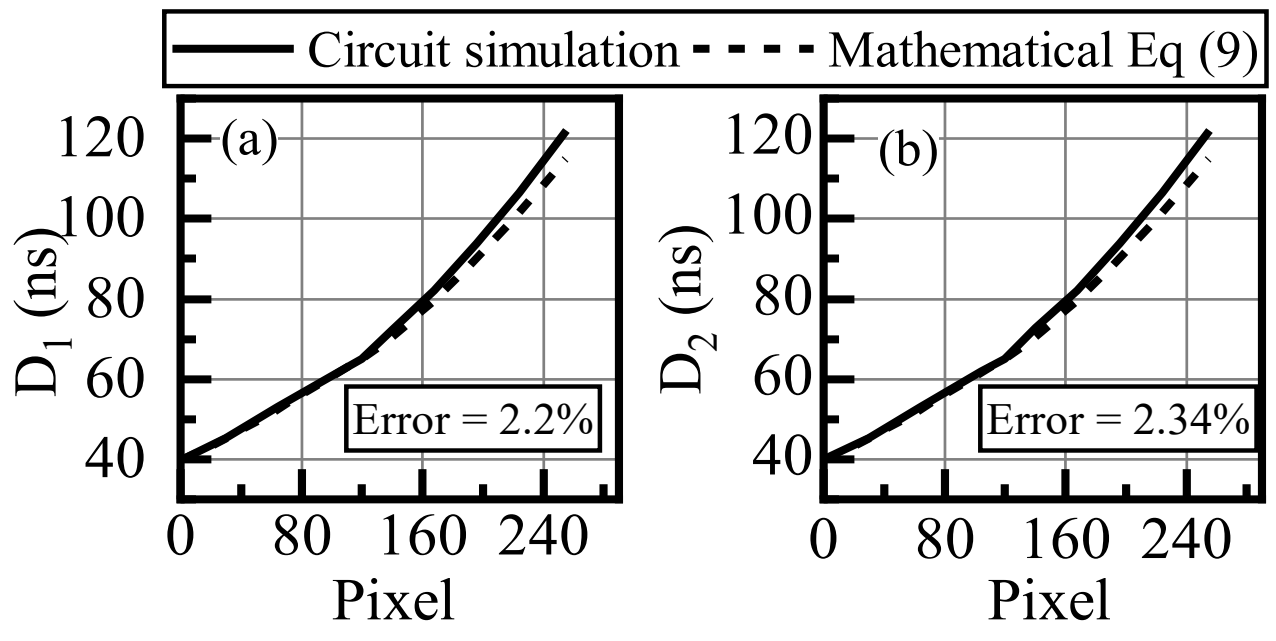

Fig. 6. Inter-spike intervals (a) D1 and (b) D2 with variation in pixel for circuit simulation and mathematical equation (9)

Fig. 6 shows the inter-spike interval ($D_1$ and $D_2$) with varying input pixel values for both the circuit simulation and the mathematical equation (9). Image from the MNIST dataset [11] has been used to analyze the encoder output. Circuit simulations are validated with the mathematical model and are found to deviate only by 2.3 %.

TABLE I. Power Consumed by single Neuron for Pixel

| Pixels | 0 | 127 | 255 |
|---|---|---|---|
| Power consumed by a single neuron (nW) | 701.57 | 543.9 | 392.1 |

Table 1 lists the power consumed by a single neuron for a given particular pixel intensity. The maximum value of 701.57nW is obtained at pixel = 0. The power decreases as we progress from black (0) to white (255) due to the neuron's decreasing excitatory current.

## IV. Conclusion

This paper presents a mathematical model and the design of a temporal neuromorphic encoder for image recognition. Image pixels are successively processed using a neuron enable signal ($V_{N_EN}$), which acts as a sampling clock. The pixels are mapped to input voltage $V_{in}$ so that conversion of images into spike trains is interpretable. A trade-off between the resolution of mapping and the speed of encoding is also established. Thus, the resolution of the image to be encoded is optimized in accordance with the speed of encoding. The mathematical model establishes an exponential relation between the inter-spike interval and the pixels of an image. The design of the temporal neuromorphic encoder is implemented with neuron enable ($V_{N_EN}$). The circuit simulation results deviate from the mathematical model by a mere 2.3% for each image in the MNIST dataset. Furthermore, the encoder's power consumption decreases with input image pixel from black (0) to white (255). This work can aid in building the mathematical view for image processing applications using a neuromorphic encoder.

## Acknowledgment

The authors are grateful to the Electrical Engineering department of Shiv Nadar University for providing simulation tools used in this work. The first author is thankful to Varsha Seshasayee and Vidisha Narang for aiding in implementing the neuromorphic encoder.